\documentclass{iau}
\usepackage{graphicx}
\usepackage{color}
\usepackage{hyperref}
\usepackage{url} 

\title[Radio source clustering towards galaxy clusters]
{Constraints on radio source clustering towards galaxy clusters:\\
application for cm-wavelength simulations of blind sky surveys. }

\author[Bartosz Lew]  
{Bartosz Lew$^1$}

\affiliation{$^1$Toru\'n Centre for Astronomy, Nicolaus Copernicus University,\\ ul. Gagarina 11, 87-100 Toru\'n, Poland\\ email: {\tt blew@astro.uni.torun.pl} }

\pubyear{2014}
\volume{xxx} 
\pagerange{119--126}
\jname{The Zeldovich Universe: Genesis and Growth of the Cosmic Web}
\editors{R. Weygaert, S. Shandarin, E. Saar and J. Einasto}

\begin{document}

\maketitle

\begin{abstract}
We constrain radio source clustering towards {\it Planck}-selected
galaxy clusters using the NVSS point source catalogue.  The constraint
can be utilised for generating realistic Sunyaev-Zeldovich effect
(SZE) mocks, and for predicting detectable clusters count and
quantifying source confusion in radio surveys.
\end{abstract}

\firstsection

\section{Introduction}
Galaxy clusters are becoming observationally useful probes of
cosmological models.  In future, thousands of galaxy clusters will be
detected via SZE.  Radio sources tend to correlate with galaxy
clusters directions and constitute an important SZE contamination. The
degree of the correlation and its redshift dependence remain uncertain
due to poor statistics of faint radio source populations and insofar
small SZE galaxy cluster samples.  In \cite{Coble2007} a constraint on
the source clustering was derived using the 28.5-GHz survey yielding
the radio source over-abundance towards galaxy cluster centres a
factor of 8.9 higher as compared to the clusters outskirts.

In preparation for the cm-wavelength sky surveys planned for
RT32/OCRA-f (One Centimetre Receiver Array installed on the 32-m radio
telescope in Toru\'n, Poland) we have calculated a possible scientific
output from such a blind survey in terms of the number of detectable
point sources and SZE-detectable galaxy clusters using hydrodynamic
simulations of large scale structure formation (\cite{Lew2015}).  In
that work we used the early Planck cluster sample (henceforth ESZ)
(\cite{PlanckCollaboration2011}) and we have shown that the point
source clustering properties may significantly alter SZE counts
predictions depending on the telescope beamwidth and observing
frequency.  In this report we revisit the point source clustering
properties towards galaxy clusters using the extended cluster sample
(henceforth PSZ) described in \cite{PlanckCollaboration2013}.

\section{Point source clustering from NVSS and Planck}
In \cite{Lew2015} we introduced a simple statistic to quantify the
clustering of radio sources towards galaxy clusters as a function of
angular distance $\theta_{\mathrm{max}}$ from cluster centres:
\begin{equation}
\rho_N(\theta_{\mathrm{max}}) = 
\frac{1}{\pi \theta_{\mathrm{max}}^2 N_0} \int_0^{\theta_{\mathrm{max}}} \frac{\partial N(\theta)}{\partial \theta} \mathrm{d}\theta \approx \frac{1}{\pi \theta_{\mathrm{max}}^2 N_0} \sum_i A_1(\theta_i)
\label{eq:rhoNtheta}
\end{equation} 
where $\rho_N(\theta_{\mathrm{max}})$ is the cumulative solid-angle
source number density, $N_0$ is the total number of clusters in the
sample and $A_1(\theta_i)=1$ if the radio source is within the angular
distance $\theta_{\mathrm{max}}$ from its associated cluster's centre
and $A_1=0$ otherwise.  The summation extends over all radio sources.

\begin{figure}[!hbt]
\begin{center}
\includegraphics[width=0.4\textwidth]{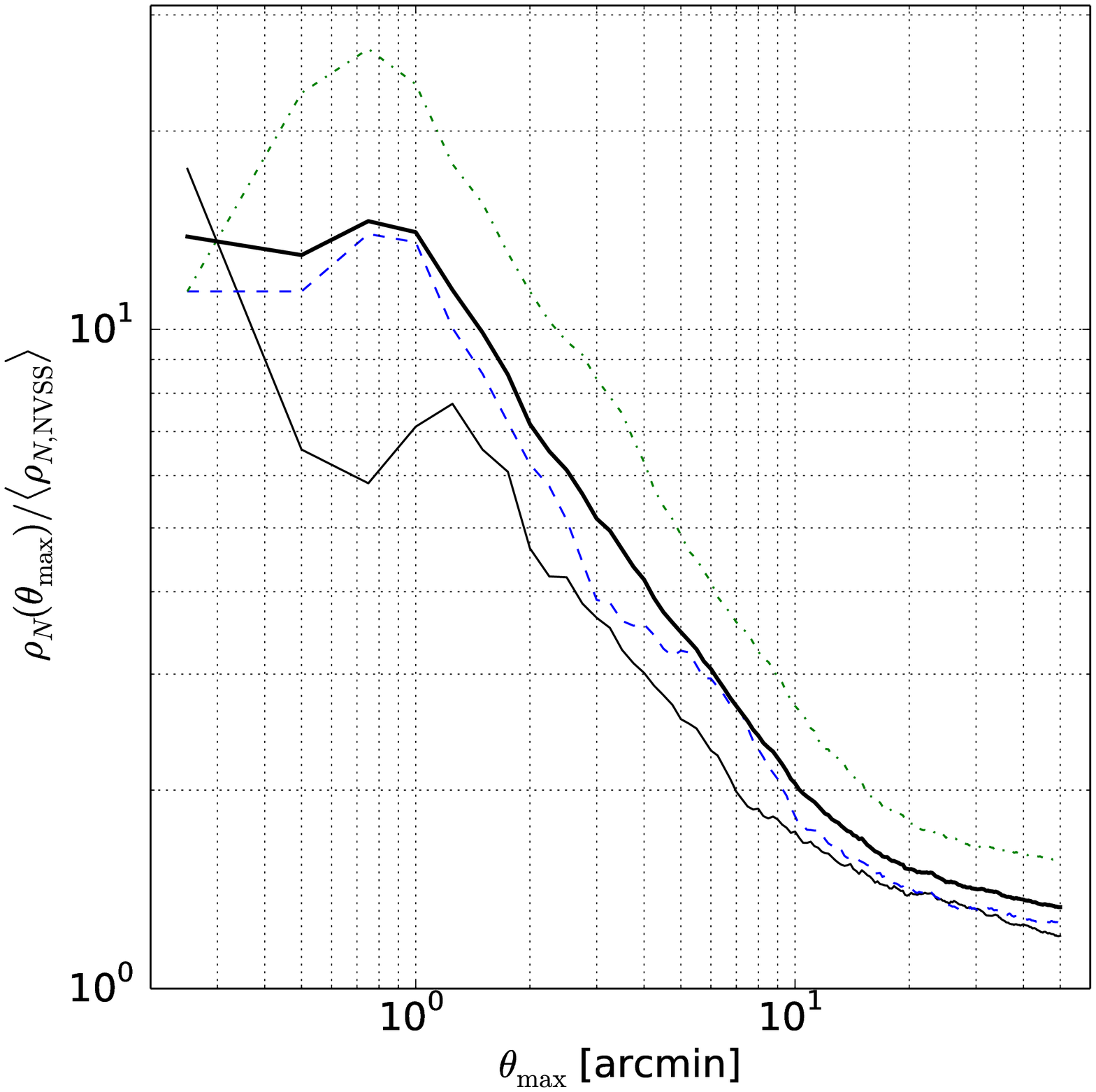} 
\includegraphics[width=0.4\textwidth]{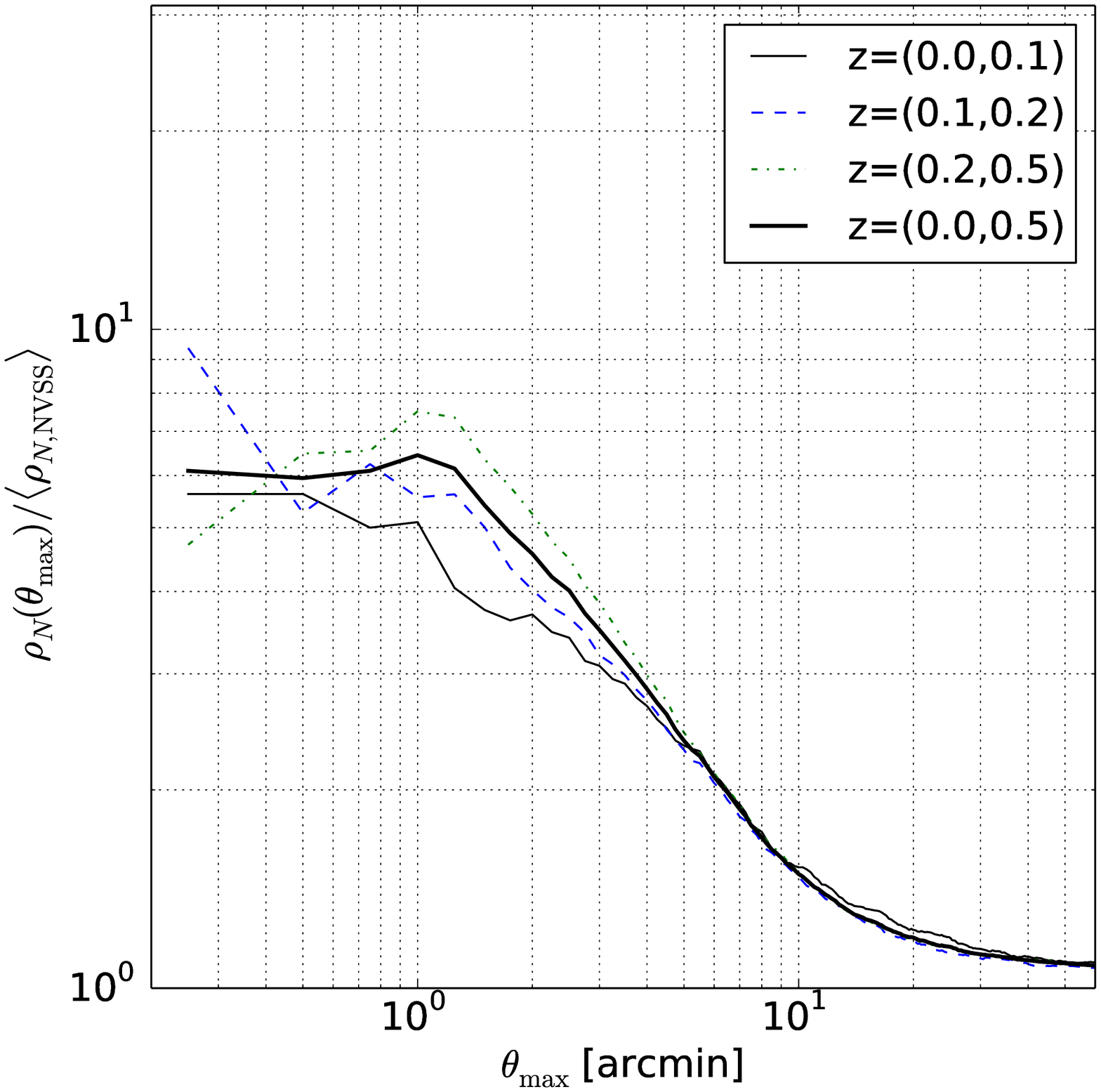} 
\caption{NVSS-normalised point source overdensity in ESZ ({\it left})
  and PSZ ({\it right}) samples as a function of angular distance from
  the cluster centres.  Individual lines represent overdensities for
  sub-samples selected according to the clusters redshift ranges.
}
\label{overdensity}
\end{center}
\end{figure}

In this report we apply this statistic to the ESZ and PSZ samples
which we cross-correlate with the 1.4 GHz NVSS radio source catalogue
(\cite{Condon1998}) out to $60'$ form the cluster centres.  In this
calculation 142 (993) out of 189 (1227) clusters were used from the
ESZ (PSZ) sample.  In Fig.~\ref{overdensity} we plot the result
relative to the NVSS average source density:
$\langle\rho_{N,\mathrm{NVSS}}\rangle \approx
0.0135\,\mathrm{arcmin}^{-2}$ which is the total number of sources in
the catalogue divided by the total survey area, and where $N_0=1$.

The point source overdensity relative to the cluster outskirts is
clearly detected in the PSZ sample as well, but its peak value is
systematically lower from the values inferred from ESZ sample in all
redshift bins.  It ranges between 5 and 10 within the innermost
$1'$. The lower values should be expected given that PSZ contains
clusters yielding lower SZE flux densities, having lower masses and
hence a lower source richness.  A more detailed analysis, taking
account of redshift space selections of radio sources (with redshifts
determined through e.g. optical identification) is needed to derive
mass-richness scaling relations, while future wide-area cm-wavelength
surveys are required to understand a plausible spectral dependence of
the clustering.  The clustering constraints are useful in making mock
surveys more realistic, especially when calculating survey confusion
limits and the expected SZE cluster count above given sensitivity
limit.
\footnote{ This work was financially supported by the Polish National
  Science Centre through grant DEC-2011/03/D/ST9/03373.  A part of
  this project has made use of ``Program Oblicze{\'n} WIElkich
  Wyzwa{\'n} nauki i techniki'' (POWIEW) computational resources
  (grant 87) at the Pozna{\'n} Supercomputing and Networking Center
  (PSNC).}

\end{document}